\documentclass[pre,aps,twocolumn,superscriptaddress,longbibliography,notitlepage]{revtex4-1}
\usepackage[colorlinks,linkcolor=darkblue,citecolor=darkblue,urlcolor=darkblue]{hyperref}
\usepackage{amsmath}
\usepackage{amsfonts}
\usepackage{xcolor}
\usepackage{verbatim}
\usepackage{graphicx}
\usepackage{esint}
\usepackage{float}
\usepackage{amssymb}
\usepackage{mathrsfs}
\usepackage{bm}
\usepackage{bbm}
\definecolor{darkblue}{rgb}{0,0,0.6}
\usepackage{circledsteps}
\usepackage{tikz,standalone}
\usetikzlibrary{matrix,positioning}
\usepackage{physics}
\usepackage{mathtools}
\newcommand{\beq}{\begin{equation}}
\newcommand{\eeq}{\end{equation}}

\usepackage{soul}
\setul{0.3ex}{0.2ex}

\newcommand{\rev}[1]{\textcolor{black}{#1}}

\begin{document}

\title{Glass and jamming transitions in a random organization model}

\author{Leonardo Galliano}

\affiliation{Dipartimento di Fisica, Universit\`a di Trieste, Strada Costiera 11, 34151, Trieste, Italy}

\affiliation{Laboratoire de Physique de l’École normale supérieure ENS, Université PSL, CNRS, Sorbonne Université,
Université de Paris, 75005 Paris, France}

\author{Ludovic Berthier}

\affiliation{Gulliver, CNRS UMR 7083, ESPCI Paris, PSL Research University, 75005 Paris, France}

\date{\today}

\begin{abstract}
We study a two-dimensional, off-lattice particle model introduced to describe absorbing phase transitions in driven non-Brownian suspensions. We numerically explore the $(\phi,\epsilon)$ phase diagram, where $\phi$ is the packing fraction and $\epsilon$ controls the amplitude of particle jumps. We use a binary mixture to suppress crystallization, which allows us to disentangle the model’s distinct phase transitions between amorphous states. At large $\phi$, we find that the approach to the absorbing transition is preceded by a non-equilibrium glass transition to a non-diffusive amorphous state. This dynamic arrest makes the location of the \rev{critical absorbing transition protocol-dependent.} The $\epsilon \to 0$ end-point of the transition line defines a jamming transition whose location is shown to vary continuously with the preparation protocol, and cannot serve as a unique definition of random close packing. Near jamming, we observe a complex landscape and marginal stability, reminiscent of Gardner phases found in thermal glasses. The critical exponents characterizing packings at the jamming transition numerically agree with alternative approaches based on energy minimization, and with analytic predictions from mean-field replica theory. We analyze \rev{hyperuniformity of large-scale density fluctuations} in fluid and glass phases, where it emerges with qualitatively distinct signatures, and show that random organization dynamics does not determine the hyperuniformity observed in jammed packings, which is found to be non-universal. Our results show that random organization models share deep physical similarities with thermal soft-particle systems undergoing glass and jamming transitions, with little impact of the non-equilibrium nature of the microscopic dynamics on emerging physical properties.    
\end{abstract}

\maketitle

\section{Introduction}

\label{sec:introduction}

Random organization models~\cite{corte2008random}, initially introduced to describe a specific experiment performed on a driven non-Brownian suspension~\cite{pine2005chaos}, now form a broad class of off-lattice particle models~\cite{corte2008random,milz2013connecting,tjhung2015hyperuniform,tjhung2016criticality,hexner2015hyperuniformity,hexner2017noise,lei2019hydrodynamics,jocteur2025random}. These models are endowed with simple local rules for particle motion that do not follow any equilibrium microscopic law, and thus drive the dynamics far from equilibrium. The typical rule is to \rev{displace overlapping particles at each time step by some small random amount.} As a result, dense systems reach a dynamical steady state where particles are \rev{repeatedly displaced,} whereas low-density ones may enter an absorbing state with no motion. These models display rich physics, such as continuous phase transitions between active and absorbing states, related to the conserved directed percolation universality class~\cite{menon2009universality,lubeck2004universal}, correlated particle dynamics in the active phases~\cite{tjhung2016criticality}, and non-trivial structural correlations with no equilibrium counterpart~\cite{hexner2015hyperuniformity}. They explore in a non-ergodic manner their configuration space~\cite{schrenk2015evidence}, frequently visiting configurations that display distinct non-equilibrium features, such as hyperuniform density fluctuations~\cite{hexner2015hyperuniformity,lei2019hydrodynamics,weijs2015emergent,ma2019hyperuniformity,Maire2025dynamical} and ordered phases in low spatial dimensions~\cite{activeXtal2023}, which are forbidden in thermodynamic equilibrium. 

A variation of the random organization model introduced biased \rev{displacements,} so that pairs of colliding particles tend to move away from each other~\cite{milz2013connecting,hexner2015hyperuniformity}. When the typical amplitude of these kicks is reduced, the biased moves act physically as a soft repulsion between particles, and the dynamics becomes close to a noisy conjugate descent in a Hamiltonian composed of pairwise repulsive forces~\cite{zhang2024absorbing,anand2026emergent}. In that case, the absorbing phase transition is shifted to considerably larger densities where particle crowding and non-equilibrium dynamics may start to compete in an interesting way~\cite{wilken2021random}. This can lead, in particular, to non-equilibrium forms of crystallization~\cite{activeXtal2023,maire2024enhancing,guo2025active} and to glassy dynamics~\cite{wang2025anomalous}. 

An interesting emerging property of random organization dynamics is that, exactly at the absorbing critical point, \rev{long-range density fluctuations are suppressed~\cite{hexner2015hyperuniformity}, or, in other words, become hyperuniform.} The static structure factor then obeys $S(q \to 0) \sim q^\alpha$, with $\alpha$ a universal exponent characterizing the conserved directed percolation universality class~\cite{wiese2024hyperuniformity}. A different form of hyperuniformity emerges in the entire active phase when the \rev{random displacements} are further constrained to conserve the position of the center of mass upon each pairwise collision~\cite{hexner2017noise}. In that case, one finds $S(q \to 0) \sim q^2$, with no need to fine-tune parameters exactly at criticality.   

When crystallization is avoided, the system remains amorphous and the absorbing transition line terminates, in the limit of vanishing kicking amplitude, at a packing fraction close to the one of the jamming transition observed in athermal packings of repulsive particles~\cite{ohern2002random,ohern2003jamming}. This has led Wilken and coworkers~\cite{wilken2021random} to propose the use of random organization dynamics as a novel definition of random close packing. The unexpected connection between random organization and jamming further led the same authors to challenge the currently-accepted value ($d_l=2$) of the lower critical dimension of the jamming transition, and to propose instead that this is controlled by conserved directed percolation for which $d_l=4$~\cite{wilken2023dynamical}. They also suggested that the hyperuniformity associated to conserved directed percolation controls density fluctuations and thus hyperuniformity at jamming. These conjectures, supported by numerical results~\cite{wilken2021random,wilken2023dynamical}, are however difficult to reconcile with the existing literature regarding the definition, location, and critical properties of the jamming transition~\cite{ohern2002random,ohern2003jamming,mari2009jamming,berthier2009glass,chaudhuri2010jamming,liu2010jamming,wyart2012marginal,charbonneau2012universal,charbonneau2015jamming,charbonneau2014fractal}.  

In this work, we numerically study a binary mixture of particles endowed with random organization dynamics, using biased particle displacements to approach large densities while remaining fully disordered. We also impose center-of-mass conservation, so that hyperuniform behavior can be expected in a large part of the phase diagram. This setting allows us to perform a detailed analysis of the structure and dynamics of the model in the various phases, as well as characterize the phase transitions between them. Our approach is therefore ideally suited to simultaneously explore the glass and jamming physics due to particle crowding, alongside the physics of absorbing phase transitions and the emergence of hyperuniformity. 

Our analysis clearly disentangle physical properties and emerging behaviors that are affected or controlled by the random organization dynamics and the conserved directed percolation universality class, from the ones that stem instead from particle crowding that typically leads to glass and jammed phases. \rev{We find that it is important to distinguish the three types of criticality that emerge in this problem: the absorbing phase transition, the jamming universality for the pair correlation singularities, and large-scale hyperuniformity emerging in distinct forms.} A central result is that, very much like in thermal systems of repulsive particles~\cite{jacquin2011microscopic}, the jamming transition is in fact buried deep inside a kinetically-arrested glass phase. In that phase, the system can be active in the random organization sense, but particles are unable to diffuse at long times. This kinetic arrest has important consequences, as it implies that the system always retains memory of its preparation at large density. This observation directly implies the existence of a continuous line of jamming transition densities, as in thermal soft spheres (the so-called $J$-line~\cite{mari2009jamming,berthier2009glass,chaudhuri2010jamming}), and therefore the impossibility to use random organization dynamics to uniquely define a random close packing transition point. Along the line of jamming transitions, we demonstrate that the critical exponents of jamming agree, to an excellent precision, with those of energy-minimized sphere packings, thus resolving an earlier controversy. We also demonstrate that hyperuniformity manifests itself differently in the liquid, glass and jammed states. We show in particular that the hyperuniformity exponent at the jamming transition is non-universal, and is thus unrelated to that of conserved directed percolation. Our work offers a complete and coherent view of glass and jamming transitions in models of random organization.

The manuscript is organized as follows. In Sec.~\ref{sec:model} we introduce the model and explain our numerical methods. In Sec.~\ref{sec:glass} we summarize the phase diagram of the model separating absorbing, glass and liquid phases. In Sec.~\ref{sec:jline} we demonstrate that the location of the jamming transition is protocol-dependent. We explore Gardner physics in the vicinity of jamming in Sec.~\ref{sec:gardner}. We measure the jamming exponents characterizing jammed packings in Sec.~\ref{sec:jamming}. The emergence of hyperunifom density fluctuations in liquid, glass and jammed configurations is separately discussed in Sec.~\ref{sec:hyperuniformity}. We summarize our results and conclude in Sec.~\ref{sec:conclusion}.

\section{Two-dimensional random organization model}

\label{sec:model}

We consider a two-dimensional model of $N$ particles in a square box of linear size $L$, with periodic boundary conditions. To suppress crystallization, we employ a bidisperse mixture where $65\%$ of the particles (type $A$) have diameter $\sigma$, while the remaining $35\%$ (type $B$) have a diameter $1.4 \sigma$. The 1.4 size ratio and 65:35 composition are known to efficiently suppress local and global ordering in two-dimensional glass models. 

\begin{figure}
    \includegraphics[width=1.\linewidth]{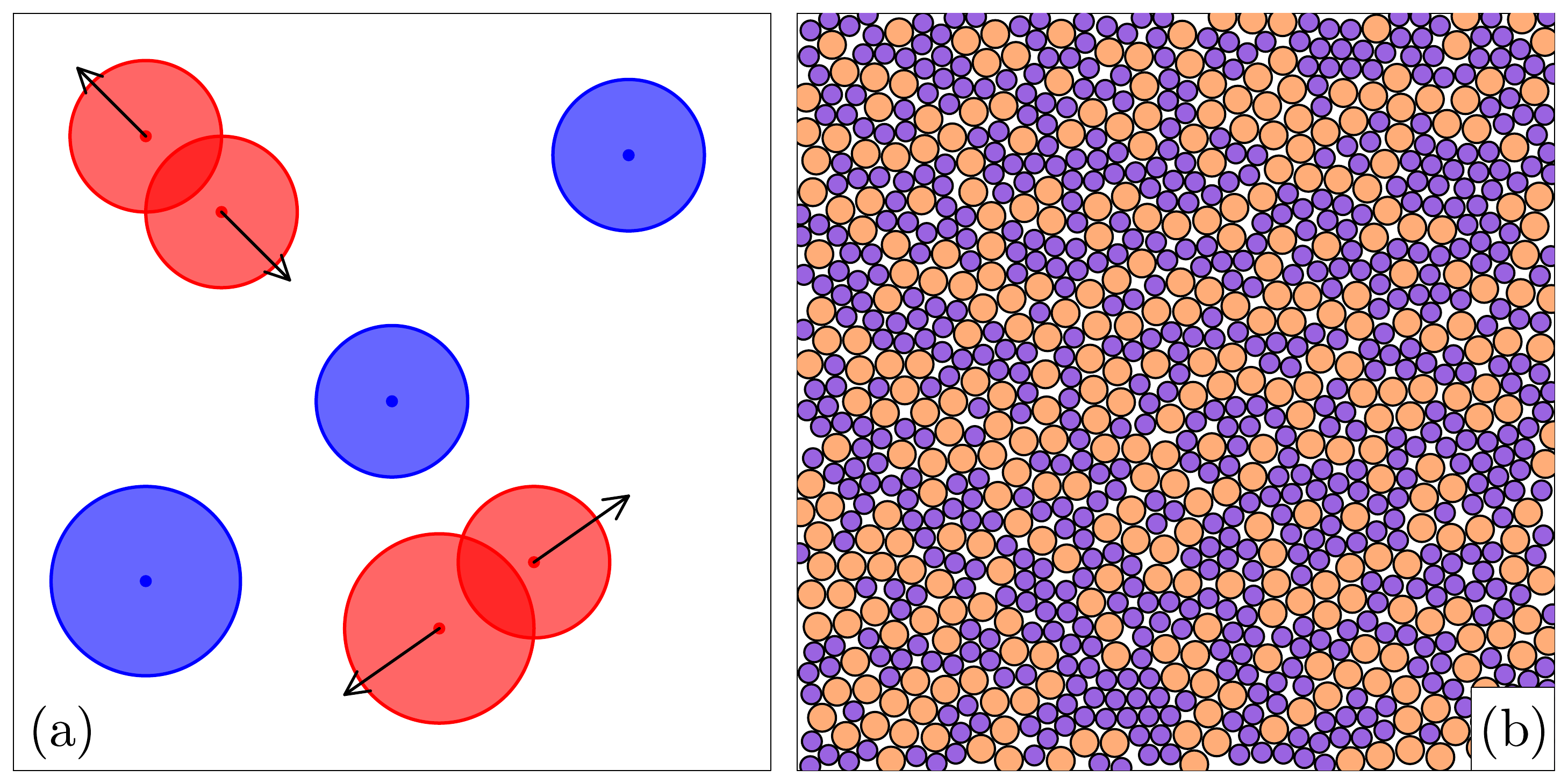}
    \caption{\rev{(a) Microscopic rules of the biased random organization model: overlapping particles (red) are given a repulsive kick of equal amplitude in opposite directions. Non-overlapping particles are inactive (blue). (b) A configuration of bidisperse disks prepared using random organization dynamics very close to the jamming transition for $N=1000$, $\phi=0.8437$ and $\epsilon=10^{-11}$.}}
    \label{fig:sketch}
\end{figure}

At each discrete time step $t$, all pairs of overlapping particles are displaced by a random amount along the line connecting their centers. Two overlapping particles $(i,j)$ are displaced in opposite directions along vectors $\vec\delta$ and $-\vec\delta$, with an amplitude $|\vec\delta|$ drawn uniformly from the interval $[0, \epsilon]$. If a particle overlaps with multiple neighbors, its total displacement at time $t$ is given by the vector sum of the individual pairwise contributions. \rev{These kinetic rules are sketched in Fig.~\ref{fig:sketch}(a).} This dynamical rule obviously conserves the position of the center of mass of the system~\cite{hexner2017noise, activeXtal2023}, as it is preserved for each individual collishansen mcdonald simple liquids bookion.

The two control parameters of the model are the packing fraction, 
\begin{equation}
\phi = \frac{\pi}{4 L^2} \sum_{i=1}^N \sigma_i^2 ,
\end{equation}
and the maximal displacement amplitude $\epsilon$. As in other random organization models~\cite{corte2008random,tjhung2015hyperuniform, hexner2015hyperuniformity, wilken2021random}, the system reaches an absorbing state at low values \rev{of either $\phi$ or $\epsilon$,} where dynamics are fully arrested and no collision occurs. By contrast, the dynamics never stops in the active phase of the phase diagram, and the system then reaches a non-equilibrium steady state. \rev{Because of the bias in the kinetic rules, the system self-organizes to remove overlaps between particles. At large density and very small $\epsilon$, the system is thus driven close to a jamming transition, which we shall study in detail. An example of a high-density configuration very close to jamming where particles barely touch is displayed in Fig.~\ref{fig:sketch}(b).}

We define the activity $f(t)$ as the fraction of overlapping particles at time $t$. Upon decreasing $\phi$ and/or $\epsilon$, the steady state activity decreases continuously and vanishes at the absorbing phase transition, $\phi_c=\phi_c(\epsilon$). Changing $\epsilon$ changes the location of the transition $\phi_c$, thus defining a line of absorbing phase transitions. Several independent studies have shown that the absorbing transition of random organization models belongs to the conserved directed percolation universality class~\cite{menon2009universality,tjhung2016criticality,wilken2021random,wilken2023dynamical}, at least when the density is not too large.  

\section{Glass transition in the active phase}

\label{sec:glass}

\begin{figure}
\includegraphics[width=\columnwidth,clip=true]{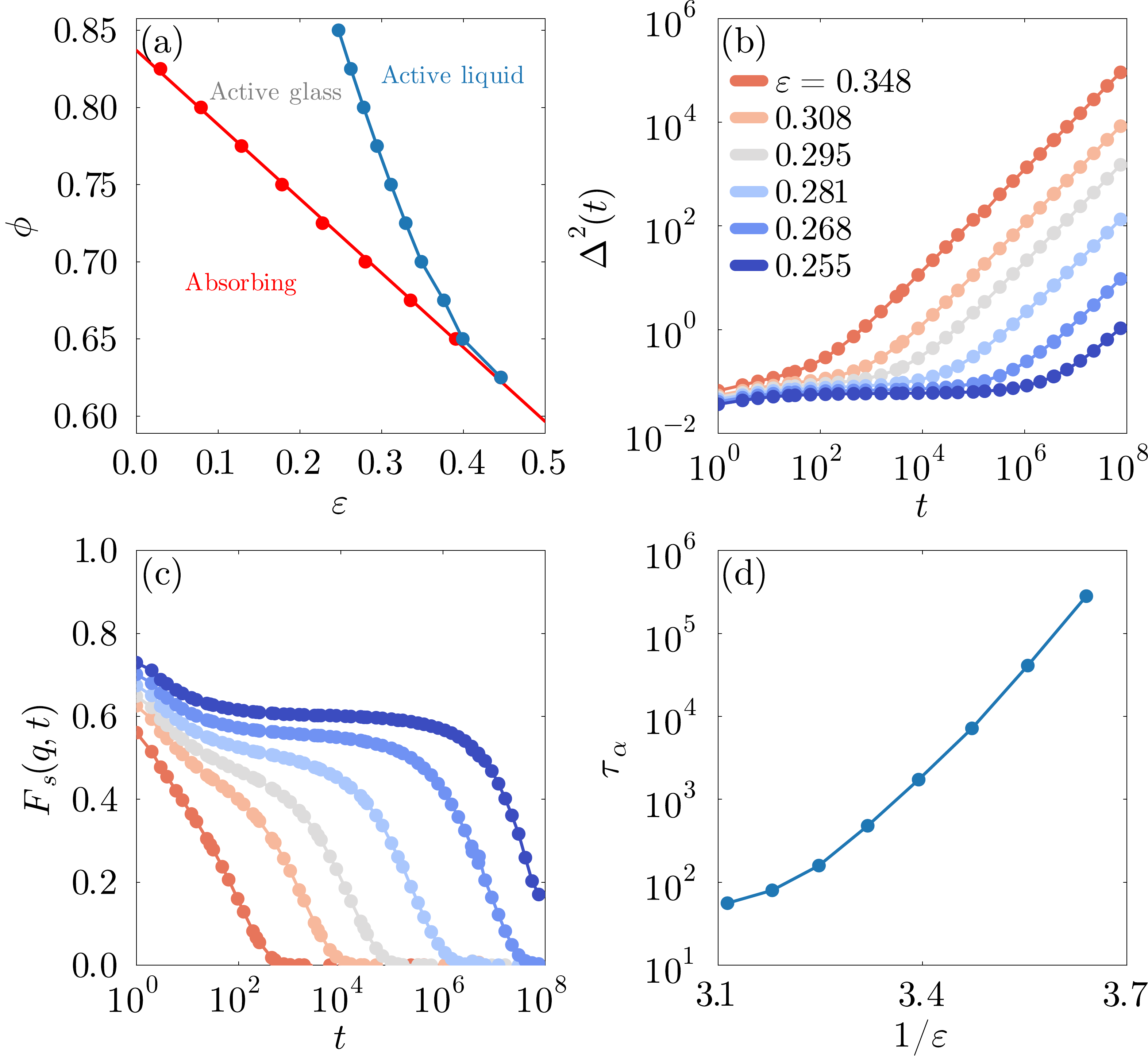} 
\caption{(a) Phase diagram showing the absorbing phase with no activity, the diffusive active liquid and the kinetically-arrested active glass.
(b) Mean-squared displacement and (c) self-intermediate scattering function at fixed $\phi=0.8$ and different $\epsilon$ revealing glassy slowing down.
(d) Corresponding structural relaxation time $\tau_\alpha$ as a function of $1/\epsilon$.}
\label{fig:glass_transition} 
\end{figure}

We first locate the phase boundary $\phi_c(\epsilon)$ separating absorbing and active phases following the conventional approach. To this end, we monitor the time evolution of the activity $f(t)$ across a wide range of state points $(\phi,\epsilon)$, starting from random initial conditions. Specifically, we vary the packing fraction across ten values in the range $\phi \in [0.65,\,0.85]$. For each $\phi$, we scan a grid of 32 values of $\epsilon \in [0,\,0.5]$. We identify the transition by distinguishing between the region where the activity $f(t)$ relaxes to zero within our observation window, and the one where it settles into a steady state with a finite activity. These simulations are performed using $N=10^3$ particles. Near the transition, finite-size fluctuations can cause the system to prematurely fall into the absorbing state, making the determination of the location of the critical line sensitive to the system size. The transition line, shown as red line in Fig.~\ref{fig:glass_transition}(a), is approximately linear in this region of the $(\phi,\epsilon)$ phase diagram. It separates the absorbing region at low $\epsilon$ from an active phase at large $\epsilon$.

We now concentrate on the active phase and quantify particle motion. We measure the mean-squared displacement
\beq
\Delta^2(t)=\frac{1}{N} \sum_{i=1}^N \langle |{\bf r}_i(t)- {\bf r}_i(0)|^2 \rangle,
\eeq
and the self-intermediate scattering function
\beq
F_s(q,t)=\frac{1}{N}\sum_{i=1}^N \langle \mathrm e^{\mathrm i {\bf q} \cdot \left( {\bf r}_i(t)- {\bf r}_i(0)\right)} \rangle,
\eeq
evaluated at the first peak of the total structure factor
\beq
S(q)=\frac{1}{N} \sum_{i,j=1}^N \langle \mathrm e^{\mathrm i {\bf  q} \cdot \left( {\bf  r}_i - {\bf r}_j \right)} \rangle .
\label{eq:soq}
\eeq
In these definitions, ${\bf r}_i(t)$ denotes the position of particle $i$ at time $t$. Due to isotropy, scattering functions only depend on $q = | {\bf q}|$. 

A central observation is that we can only measure the full decay of the self-intermediate scattering function, or the long-time diffusive regime of the mean-squared displacements in a restricted part of the phase diagram. We show representative measurements in Figs.~\ref{fig:glass_transition}(b,c) taken along a constant $\phi=0.8$ line by varying $\epsilon$.  While the dynamics is fast and diffusive for large $\epsilon$, particle motion slows down dramatically as $\epsilon$ is decreased. We extract a relation time using the definition  $F_s(q, t=\tau_\alpha) = e^{-1}$ and report its evolution with $\epsilon$ in Fig.~\ref{fig:glass_transition}(d). We use a graphical representation where $\log \tau_\alpha$ is shown against $1/\epsilon$ as it emphasizes the similarity between the present model and the physics of thermal systems undergoing a glass transition upon decreasing the temperature~\cite{berthier2011theoretical}. As $\epsilon$ controls the amplitude of particle motion, it plays a role qualitatively analogous to temperature in Brownian systems. We did not attempt a deeper connection between $\epsilon$ and any definition of an effective temperature, which could be an interesting question for future work. The two-step decay of time correlation functions, and the rapid growth of the structural relaxation time are well-known signatures of the glass transition~\cite{berthier2011theoretical}. 

There is no unique way to determine the respective location of the glass and liquid phases in the phase diagram, as the glass `transition' represents, in practice, a dynamic crossover. Yet, the sharp increase of $\tau_\alpha$ with decreasing $\epsilon$ shows that any practical definition will yield a qualitatively similar distinction between the two phases. We decided to represent the glass-liquid kinetic arrest line as the contour in the $(\phi, \epsilon)$ plane where $\tau_\alpha = 10^5$, see Fig.~\ref{fig:glass_transition}(a). For the values of $\epsilon$ shown, the steady-state activity $\langle f \rangle$ lies in the range $[0.88, 0.96]$, meaning that the system is well within the active phase and far from the absorbing transition. The observed slowing down is therefore a consequence of particle crowding and is unrelated to the critical slowing down associated with the absorbing transition itself~\cite{tjhung2016criticality}. The two transition lines only coincide at much larger values of $\epsilon$, when crowding plays a negligible role~\cite{corte2008random,tjhung2016criticality}. 

The observation that the active region at large density of the random organization model is a glass rather than a diffusive liquid implies that memory of the preparation of the system is never forgotten. Memory is indeed the most distinctive feature of glasses~\cite{berthier2016facets}. While the system can explore a large number of distinct configurations in the diffusive liquid region, the physics is very different in the glass, active region, where particle merely perform confined motion (sometimes described as a `caged' dynamics) around well-defined positions that form, at large scale, an amorphous structure. 

In the rest of the manuscript, we shall explore several important consequences of the existence of the kinetically arrested glass phase. 

\section{A continuous line of dynamic jamming transitions}

\label{sec:jline}

\subsection{Continuous family of absorbing transition lines}

The phase diagram in Fig.~\ref{fig:glass_transition}(a) shows that for $\epsilon \leq 0.3$, the absorbing phase transition at $\phi_c(\epsilon)$ occurs between an arrested (non-diffusive) glass state and an inactive phase. As the glass, almost by definition, retains the memory of its preparation, the location of the phase boundary can in principle also be history-dependent. The choice of random initial conditions in Fig.~\ref{fig:glass_transition}(a) is a totally valid one, but it is neither unique nor superior. There exists in fact an infinite set of equally valid, but distinct, initial conditions. 

To illustrate this, we design a family of initial conditions parametrized by a single control parameter. We first recall that random initial conditions can be thought of as representing equilibrium configurations of a hard disk system at infinite dilution, $\phi \to 0$, or equivalently, at zero pressure, $P \to 0$. For hard disks, it is convenient to use the reduced pressure, or compressibility factor~\cite{hansen2013theory}
\begin{equation}
Z = \frac{P}{\rho k_B T},
\end{equation}
where $\rho$ is the number density and $k_B T$ the thermal energy. 

To construct a continuous family of initial conditions, we use a finite pressure $Z_0>0$ to generate equilibrium hard disk configurations using Monte Carlo simulations in the $NPT$ ensemble~\cite{frenkel2023understanding}. At each pressure $Z_0$ we can thus easily produce an ensemble of equilibrium configurations. We use $N=10^3$ particles. These configurations are then used as initial conditions to explore any state point in the $(\phi, \epsilon)$ phase diagram of random organization, using affine compressions to achieve the desired packing fraction $\phi$.

\begin{figure}
\includegraphics[width=\columnwidth,clip=true]{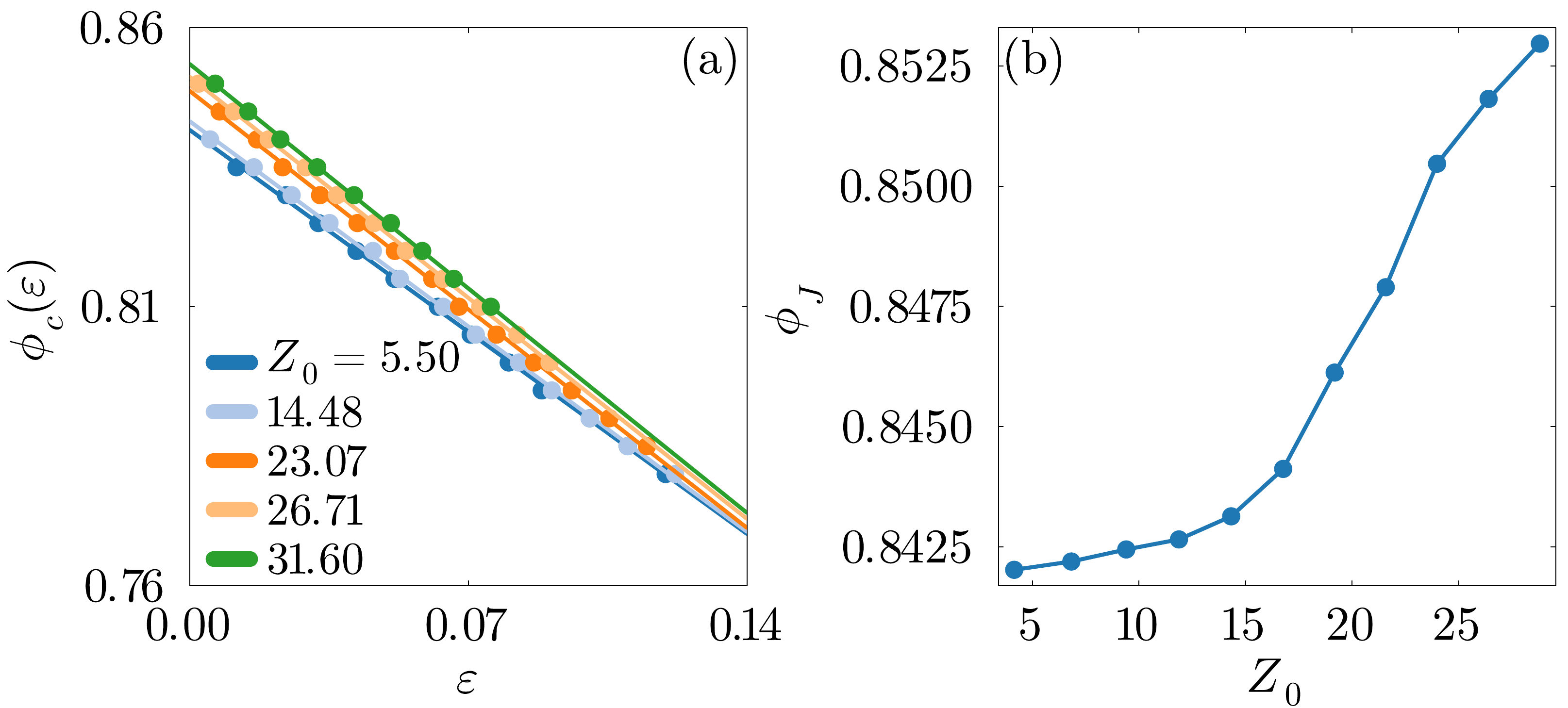} 
\caption{(a) The critical line $\phi_c(\epsilon)$ between the active glass and  absorbing states obtained using different sets of initial conditions parametrized by the reduced pressure $Z_0$ depends continuously on $Z_0$, because the active glass retains memory of its preparation. 
(b) The jamming packing fraction $\phi_J$, Eq.~(\ref{eq:phij}), obtained under random organization dynamics also depends  on $Z_0$, and is therefore not universal.}   
\label{fig:jline} 
\end{figure}

For each ensemble of configurations produced at reduced pressure $Z_0$, we determine the critical line $\phi_c(\epsilon)$ separating active and absorbing states using the same procedure as in Sec.~\ref{sec:glass}. By construction, the absorbing transition line shown in Fig.~\ref{fig:glass_transition}(a) corresponds here to the $Z_0 \to 0$ limit. The results in Fig.~\ref{fig:jline}(a) show that the location of the absorbing transition line in the $(\phi,\epsilon)$ phase diagram depends continuously on the preparation protocol, via the value of $Z_0$. We observe that the critical packing fraction $\phi_c(\epsilon)$ for systems prepared at larger $Z_0$ values increases with $Z_0$ at constant $\epsilon$. Note that when $\epsilon$ becomes large, the dependence on $Z_0$ disappears as the active phase is now a diffusive liquid, that retains no memory of its initial preparation.

In the small $\epsilon$ part of the phase diagram, where the active phase is not diffusive, there exists no protocol that can fully erase the memory of the initial conditions, and the absorbing transition line necessarily depends continuously on all parameters involved in the preparation protocol.   

\subsection{Continuous evolution of jamming density: $J$-line}

As the transition line $\phi_c(\epsilon)$ depends continuously on all parameters of the chosen protocol for small $\epsilon$, this is also true of its limit $\epsilon \to 0$. In this limit, the critical configurations produced under random organization dynamics have been shown~\cite{wilken2021random} to be similar to packings of soft repulsive particles prepared exactly at jamming using energy minimization~\cite{ohern2003jamming}. Wilken and coworkers thus proposed
\begin{equation}
  \phi_J = \phi_c(\epsilon \to 0),
  \label{eq:phij}
\end{equation}
as a unique definition of random close packing~\cite{wilken2021random}. In practice, the definition (\ref{eq:phij}) should be understood as a double limit, where the packing fraction $\phi$ needs to approach the absorbing transition line, $\phi \to \phi_c$ for continuously decreasing values of $\epsilon \to 0$.

Here, we argue instead that this identification cannot be unique and that random organization dynamics in fact produces a line of critical jamming densities. A similar conclusion in fact holds for the `$J$-point' of energy-minimized packings~\cite{ohern2002random}, that was shown long ago to be better qualified as a `$J$-line'~\cite{parisi2005ideal,mari2009jamming,chaudhuri2010jamming}. The same conclusion is also true for hard sphere compression algorithms~\cite{berthier2009glass}. This is presumably unavoidable~\cite{donev2004comment}: any preparation algorithm to get to jamming must displace and carefully rearrange particles, but the system is so crowded and constrained that jamming necessarily occurs deep within a glassy phase, be it thermal or not. The broad conclusion~\cite{chaudhuri2010jamming} is that what matters is perhaps not the precise location of the jamming transition, which is non-universal, but the critical exponents that describe physical behavior in its vicinity, which are believed to be universal (see Sec.~\ref{sec:jamming} below).  

To support our hypothesis, we need to design a numerical procedure to efficiently approach the $\phi_c(\epsilon \to 0)$ limit. We mimic the procedure used for soft repulsive particles where the density is gradually (or sometimes, iteratively~\cite{desmond2009random}) adjusted to be as close as possible to the jamming transition~\cite{ohern2002random,charbonneau2015jamming}, but here we of course need to use random organization dynamics only.  

In practice, we begin by quenching equilibrium hard-disk configurations of $N=5 \times 10^3$ particles prepared at a given equilibrium pressure $Z_0$ to the state point $(\phi=0.82, \epsilon=0.01)$, which ensures that all configurations initially contain overlaps. We then implement an annealing procedure, where $\epsilon$ is slowly decreased at each step of our dynamics using an exponential decay: $\epsilon(t+1) = \epsilon(t) (1-\lambda)$, with $\lambda>0$ a parameter that controls the rate of decrease of the variable $\epsilon$. If $\epsilon$ were decreased at constant $\phi$ the system would end up in the absorbing phase, while we wish to remain on the critical line. To this end, we monitor the activity $f(t)$ during the  annealing, and adjust the packing fraction $\phi$ to maintain $f \in [0, 0.2]$. When $f$ exceeds (falls below) this window, we iteratively decrease (increase) $\phi$ until the activity falls back into the desired range. We stop the annealing protocol after a total time $T$. If the system is still active at time $T$, we keep annealing $\epsilon$ at constant $\phi$ to obtain an absorbing configuration with no overlap. This packing fraction constitutes our best numerical estimate of $\phi_J$ defined in Eq.~(\ref{eq:phij}) as the mathematical $\epsilon \to 0$ limit. For each set of initial conditions, the value of $\phi_J$ is averaged over several realizations of the annealing protocol, using up to 4 independent initial configurations and 10 annealing protocols per configuration. 

This numerical protocol mainly depends on two control parameters: the annealing rate $\lambda$ and the stopping time $T$. Varying them allows us to approach the limit $\phi_c(\epsilon \to 0)$ with an arbitrary precision. We shall see below in Sec.~\ref{sec:jamming} that perfectly isostatic packings require careful preparation with $\lambda \ll 1$. Here, we simply locate the jamming transition point $\phi_c(\epsilon \to 0)$ and its evolution with initial conditions parametrized by $Z_0$. We find that the values $\lambda = 1.8 \times 10^{-7}$ and $T = 5 \times 10^7$ are sufficient, as we can reach values $\epsilon \approx 10^{-6}$ at time $T$ while being very close to criticality at $\phi_c$, within a reasonable amount of CPU time. 

We report in Fig.~\ref{fig:jline}(b) the continuous evolution of the jamming packing fraction $\phi_J$ as a function of the pressure $Z_0$ characterizing the initial condition. As anticipated, we find that $\phi_J$ increases monotonically with $Z_0$. These results demonstrate that there is no unique jamming point under random organization dynamics, but instead a line of jamming transitions that depends on the preparation history of the system. These results echo similar results obtained using hard sphere compressions~\cite{berthier2009glass}, or energy-minimization protocols using soft particles~\cite{chaudhuri2010jamming,ozawa2017exploring,berthier2024Monte}. In particular the choice $Z_0=0$ (random initial conditions) is close in spirit to the early definition of the $J$-point by O'Hern and coworkers~\cite{ohern2002random}. This similarity is not surprising as random organization dynamics was analytically shown to become analogous to a noisy gradient descent dynamics in a soft repulsive potential when $\epsilon \to 0$~\cite{zhang2024absorbing,anand2026emergent}, which is indeed very close to the protocol proposed by O'Hern {\it et al.}.     

The broader conclusion is thus the confirmation that the location of the jamming transition is not universal, and depends on all details of the protocol used to reach jamming~\cite{donev2004comment}. Our results show that this known conclusion applies to random organization dynamics. From a physics perspective, this observation considerably undermines efforts to define or predict quantitatively the actual value of $\phi_J$. These efforts in fact only make sense when all details of the specific protocol used to arrive to jamming are included in the prediction, which seems unrealistic. 

In the rest of the paper we focus separately on physical behavior and quantities that are universal, and the ones that are protocol-dependent. 

\section{Emergence of Gardner physics near jamming}

\label{sec:gardner}

A remarkable theoretical discovery was made when replica calculations were applied to study the jamming transition~\cite{parisi2005ideal,parisi2010mean}. Initial attempts using one-step replica symmetry breaking made predictions for some physical quantities near jamming that disagreed with simulations~\cite{parisi2005ideal,parisi2010mean,jacquin2011microscopic,berthier2011microscopic}. This problem was then cured when the existence of a Gardner transition to a full-replica symmetry breaking glass phase was established~\cite{kurchan2013exact}. The approach to jamming was then correctly predicted by analytic calculations for all quantities~\cite{charbonneau2014exact,charbonneau2017glass,parisi2020theory}, and agreement with simulations was found to be essentially perfect in all spatial dimensions down to $d_l=2$, considered for that reason as the lower critical dimension for jamming. There is therefore a strong connection between the emergence of Gardner physics and the criticality associated to the jamming transition~\cite{berthier2019gardner}.  

Despite the absence of an energy function or underlying Hamiltonian, the present random organization model exhibits a jamming transition with associated criticality. It is thus natural to ask whether Gardner physics is also observed near jamming, by analogy with thermal particle systems~\cite{charbonneau2014fractal, berthier2016growing, scalliet2017absence,seoane2018spin,liao2019hierarchical,scalliet2019nature}. In mean-field calculations, the Gardner transition is associated with a hierarchical splitting of the energy landscape into a complex structure of marginally stable sub-basins, leading to a loss of ergodicity deep within the (already non-ergodic) glass phase~\cite{charbonneau2014fractal}. \rev{Technically the Gardner transition is detected 
by searching for an instability in the evolution of the thermodynamic free energy. In computer simulations, it is instead detected using purely dynamic observables related to small-scale motion of the particles near their time-averaged positions. This strategy paved the way for similar analyses in dense particle systems that evolve far from equilibrium~\cite{seguin2016experimental,xiao2022probing}. Here we ask similarly whether, given the complete absence of an energy function and therefore of a free energy landscape, dynamic signatures of Gardner physics can nevertheless be detected.} 

To answer this question, we employ a protocol similar to those used to detect Gardner physics in thermal glasses~\cite{berthier2016growing,scalliet2019nature}. We first prepare 9 independent glass configurations of $N=10^3$ particles at $(\epsilon_0 = 0.1, \phi_0 = 0.8368)$ starting from random initial conditions and we allow them to evolve for a fixed waiting time $T=10^8$. This state point is deep in the active glass part of the phase diagram. After this long waiting time, we have an ensemble of well-aged glass configurations that are used to further probe the complexity of the configuration space within the glass phase. 

For each configuration, we generate 4 identical copies (also called clones, or replica) and instantaneously quench them to $\phi=0.84$ and various values of $\epsilon$, running random organization dynamics with independent realizations of the stochastic noise.

The expectation is that if the configuration space of the glass phase is composed of a simple basin, different copies should fully explore that basin in such a way that the distance each copy can travel is equal to the distance between them (think of the clones as non-interacting particles in a single harmonic well). Instead, if the configuration space is fractured in a large number of basins separated by large barriers, independent copies will only explore a restricted part of the configuration space. In that case the distance between copies is possibly much larger than the distance each copy can travel. 

\begin{figure}
\includegraphics[width=\columnwidth,clip=true]{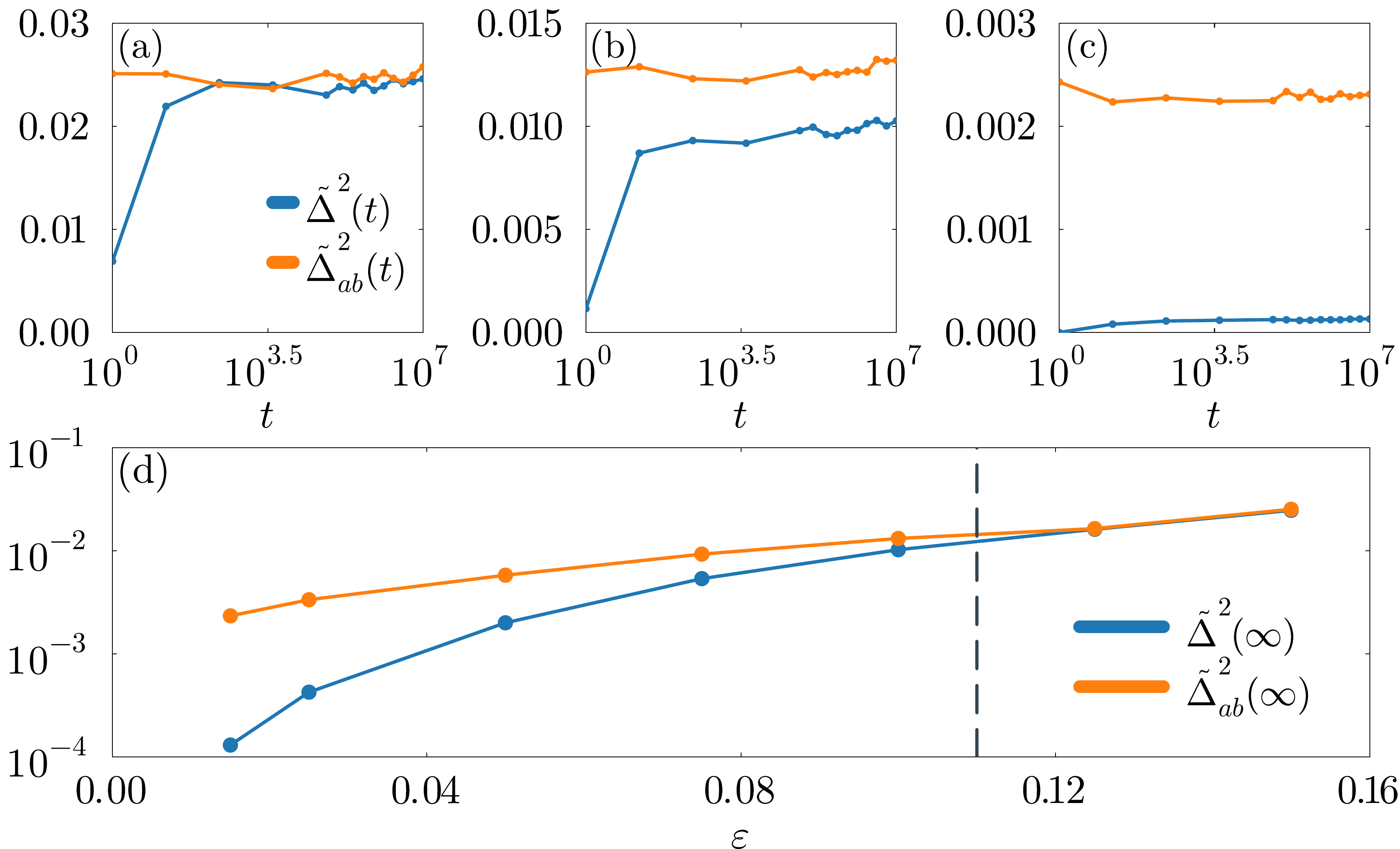} 
\caption{(a-c) Time evolution of the mean-squared distances  $\tilde{\Delta}^2(t)$ from Eq.~(\ref{eq:msdgardner}) and $\tilde{\Delta}^2_{ab}(t)$ from Eq.~(\ref{eq:msdabgardner}) for (a) $\epsilon=0.15$, (b) $\epsilon=0.1$, and (c) $\epsilon=0.015$.
The distance between copies becomes much larger than the distance each copy travels at low $\epsilon$, indicating ergodicity breaking.  
(d) The long-time plateau values gradually split when $\epsilon$ decreases, revealing a Gardner crossover near $\epsilon \approx 0.11$ (dashed line) and the emergence of Gardner physics on the approach to jamming at $\epsilon \to 0$.} 
\label{fig:gardner} 
\end{figure}

To distinguish between these two cases, we introduce two metrics: one for the distance traveled by each copy, and another for the distance between them. For the former we use the definition 
\beq
\tilde{\Delta}^2(t) = \frac{\pi}{N_{\text{B}}} \langle \sum_{i\in\text{B}} |{\bf r}_i(t)- {\bf r}_i(0) |^{-1} \rangle^{-2},
\label{eq:msdgardner}
\eeq
where the sum is restricted to the large particles (of type $B$). This metric is close in spirit to a mean-squared displacement (they have the same dimension) but our choice of a different moment of the distribution of particle displacements reduces the possible influence of a few very mobile particles that could otherwise dominate the conventional mean-squared displacement~\cite{ikeda2013dynamic}. It was in particular shown that such localized defects would give rise to fictitious signatures of Gardner physics even in the absence of a Garner phase transition~\cite{scalliet2017absence}, while the choice in Eq.~(\ref{eq:msdgardner}) is an easy cure to this problem. \rev{Finally the prefactor $\pi$ ensures that $\tilde{\Delta}^2$ is equal to the usual mean-squared displacement when the distribution of displacements is a Gaussian.} The brackets represent an average over all $9 \times 4 = 36$ copies.

The second metrics measures the average distance between two replica $(a)$ and $(b)$ originating from the same initial configuration:
\beq
\tilde{\Delta}_{ab}^2(t) = \frac{\pi}{N_{\text{B}}} \langle \sum_{i\in\text{B}} | {\bf r}^{\,(a)}_i(t)-{\bf  r}^{\,(b)}_i(t) |^{-1} \rangle^{-2},
\label{eq:msdabgardner}
\eeq
where ${\bf r}^{\,(a)}_i(t)$ denotes the position of particle $i$ in copy $(a)$. 

The time evolution of both observables is shown in Figs.~\ref{fig:gardner}(a–c) for three values of $\epsilon$. At large $\epsilon=0.15$, the two curves rapidly reach the same plateau value, indicating that all copies easily explore the same part of the configuration space. By lowering $\epsilon$ below $\epsilon \approx 0.11$, the two metrics no longer converge to the same value, indicating a loss of ergodicity. We measure the long-time value of both metrics, and their gradual splitting as $\epsilon$ decreases is shown in Fig.~\ref{fig:gardner}(d).

Similar signatures of an ergodicity breaking transition happening within glass phases have been reported in numerous simulations of thermal glasses before, in the context of the Gardner transition~\cite{berthier2016growing,berthier2019gardner,scalliet2019nature,liao2019hierarchical}. A fewer number of experimental studies have been performed and similar signatures of Gardner physics were obtained~\cite{seguin2016experimental,xiao2022probing}. Experimental work was mostly performed for non-equilibrium driven systems. Our numerical work using a non-thermal dynamics confirms that signatures of Gardner physics are indeed not restricted to thermal systems, but are likely a generic feature of particle systems approaching a jamming transition~\cite{berthier2019gardner,charbonneau2021memory}.

Finally, the analogous behavior observed between thermal systems and the present random organization dynamics again emphasizes the great similarity of behavior between the two approaches when it comes to the criticality surrounding the jamming transition. This also demonstrate the relevance of mean-field replica predictions and calculations to account for the physics observed in random organization models. 

\begin{figure*}
\includegraphics[width=2\columnwidth]{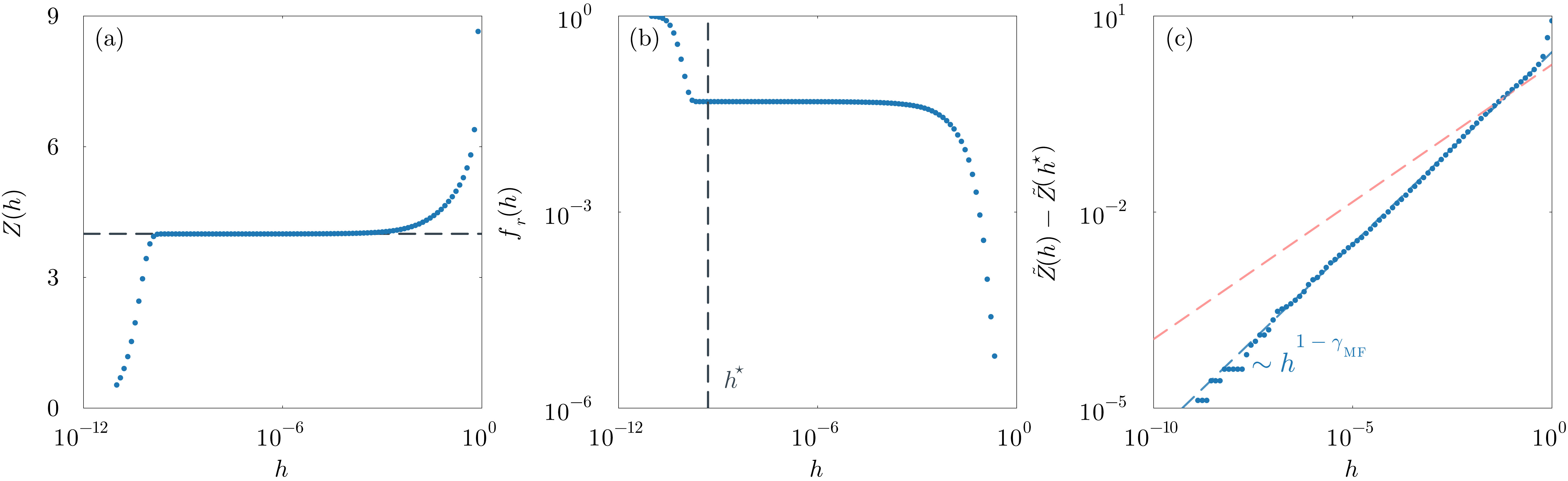} 
\caption{(a) Average number of neighbors $Z(h)$ within a gap $h$. The extended plateau close to $Z = 4$ (dashed line) reveals a close approach to isostaticity and to the jamming critical point.
(b) Evolution of the fraction of rattlers $f_r$ for different definitions $h$ of contacts. We select $h^\star = 5 \times 10^{-10}$ (dashed line) to define contacts and remove rattlers. 
(c) The excess contact number $\tilde Z(h)-\tilde Z(h^\star)$, Eq.~(\ref{eq:zh}), is well-described by the mean-field exponent $\gamma = 0.41269\cdots$ (blue), but differs from the value $\gamma=0.58$ (red) reported in Ref.~\cite{wilken2023dynamical} for a similar model.}
\label{fig:jamming} 
\end{figure*}

\section{Resolving a controversy about jamming exponents}

\label{sec:jamming}

We now consider the $\epsilon \to 0$ limit of the absorbing state critical line and focus on the jamming end point itself. Wilken and coworkers have shown that the configurations obtained in this limit share deep similarities with earlier observations made on jammed packings~\cite{wilken2021random}. In particular, isostaticity is satisfied with an averaged contact per particle $Z_{\text{iso}} = 2d$ in an infinite system in $d$ dimensions, which represents the minimal number of mechanical contacts required for mechanical rigidity. However, numerical results also indicated that one of the two independent exponents characterizing the jamming criticality differs quantitatively with literature results in dimensions $d=2, 3$~\cite{wilken2023dynamical}. They concluded that isostatic packings prepared under random organization dynamics display a distinct criticality, together with a different lower critical dimension, possibly the one ($d_l=4$) dictated by conserved directed percolation~\cite{wilken2023dynamical}. Our numerical analysis and conclusions will differ from those of Wilken {\it et al.}.

Past work has shown that to rigorously test the criticality of jammed packings, very careful preparation is needed to approach extremely closely to the critical point~\cite{charbonneau2012universal,charbonneau2015jamming}. We will show that the numerical measurements reported in Ref.~\cite{wilken2023dynamical} are taken too far from the jamming critical point, leading to erroneous conclusions and interpretations.

We prove our assertion by employing the same annealing protocol as in Sec.~\ref{sec:jline}. To reach state points as close as possible to the jamming critical point, we decrease as much as we can the annealing rate $\lambda$ and increase the maximal annealing time $T$, within the limits imposed by our computer resources. This allows us to push the annealing procedure down to values $\epsilon \approx 10^{-10}$ while maintaining the system in close vicinity to the critical line. This is about $10^7$ times smaller than the smallest $\epsilon$ values analyzed in Ref.~\cite{wilken2023dynamical}.

In practice, we use an ensemble of 32 independent configurations with $N=5 \times 10^3$ particles initially evolved at $(\epsilon_0 = 0.01, \phi_0 = 0.84)$ starting from random initial conditions and aged for $10^8$ steps, corresponding to about 100 times the number of steps required for the activity $f(t)$ to reach a steady state. We set the annealing rate to $\lambda= 0.92 \times 10^{-7}$ and the total number of steps to $T=2 \times 10^8$. 

At the end of the annealing, the value of $\epsilon$ is $\epsilon \approx 10^{-10}$, and most configurations are active with an activity near $f \approx 0.2$. We have verified that the system is then at a distance $\delta \phi \approx 10^{-11}$ from $\phi_c$ at this value of $\epsilon$. To approach the critical line, we keep annealing $\epsilon$ slowly at constant $\phi$ until the activity vanishes. This effectively removes all overlaps from the jammed configurations. We checked that by decreasing the annealing rate in this final step by several orders of magnitude does not impact the gap distribution. This shows that our protocol efficiently implements the double limit $\epsilon \to 0$ and $\phi \to \phi_c$ implied in Eq.~(\ref{eq:phij}). The crucial difference between our approach and that of Wilken {\it et al.} is the gradual annealing of $\epsilon$ towards much smaller values, which allows us to control the distance to jamming with very high precision in both $\epsilon$ and $\phi$ variables.

To illustrate the close proximity to the jamming transition, we compute the average number of neighbors within a gap $h$
\beq
Z(h)=\frac{1}{N}\sum_{j\neq i}\left<\Theta\left(h-\frac{r_{ij} - \sigma_{ij}}{\sigma_{ij}}\right)\right>,
\eeq
where $r_{ij} = | {\bf r}_i - {\bf r}_j|$, $\sigma_{ij} = (\sigma_i + \sigma_j)/2$ and $\Theta(x)$ is the Heaviside function. For truly isostatic configurations, $Z(h \to 0) = 4$ for $d=2$~\cite{charbonneau2012universal}. As shown in Fig.~\ref{fig:jamming}(a), our configurations display $Z(h) \approx 4$ over several decades in $h$ between $h=10^{-10}$ and $10^{-3}$. In particular, these data show that our approach to jamming is superior to that in Ref.~\cite{wilken2023dynamical}, where data stop below $h=10^{-3}$ and $Z(h)$ does not display a clear plateau. 

We now perform a detailed analysis of the gap distribution. We first identify and remove all rattlers, defined as particles with fewer than $d+1$ contacts that do not contribute to mechanical stability~\cite{charbonneau2015jamming}. The fraction of rattlers $f_r(h)$ can be followed as a function of the gap $h$ used to define a contact, see Fig.~\ref{fig:jamming}(b). This function exhibits an extended plateau in the same range $h \in [10^{-10}, 10^{-3}]$, with a constant value fraction of rattlers of about $5 \%$.

This observation allows us to safely determine a cutoff $h^\star = 5 \times 10^{-10}$ to meaningfully define a contact as particles with a gap small than $h^\star$. We can then iteratively remove rattlers using that contact scale, until no more particles are identified as rattlers. We define $\tilde Z(h)$ as the average number of neighbors within a gap $h$, after rattlers have been removed.

To assess the mechanical stability of our packings, we compare $\tilde Z(h^\star)$ with $Z_{\text{iso}}=2(d(N-1)+1)/N$, which is the number of contacts of a perfectly isostatic configuration with a finite number, $N$, of particles~\cite{charbonneau2015jamming}. We find that the distance to isostaticity $\Delta Z = | \tilde Z(h^\star)-Z_{\text{iso}} |$ decreases systematically as the annealing rate $\lambda$ is decreased. For the slowest protocol employed here (with $\lambda=0.92 \times 10^{-7}$) we find $\Delta Z \approx 0.003$. For a packing composed of $N=5 \times 10^3$ particles and thus of about $10^4$ contacts, we are only missing about 7 contacts on average per configuration. It would be interesting to improve even further the preparation of jammed packing to obtain perfect isostaticity, as needed to measure the critical behavior of the force distribution~\cite{charbonneau2017glass}.
 
The jamming critical exponent $\gamma$ describes the number of near-contacts in jammed packings. Using the above notations, this is defined as
\beq
\tilde Z(h) - \tilde Z(h^\star) \propto h^{1-\gamma}.
\label{eq:zh}
\eeq
The definition in Eq.~(\ref{eq:zh}) is equivalent to the divergence $g(r) \sim 1/(r-1)^\gamma$ of the pair correlation function at the contact value, $r \to 1^+$. Mean-field replica calculations predicts $\gamma \approx 0.41269 \cdots$~\cite{charbonneau2015jamming}, confirmed by several numerical studies down to $d_l=2$ using energy-minimized packings~\cite{charbonneau2017glass}. Instead, Wilken {\it et al.} report their largest discrepancy in the $d=2$ random organization model~\cite{wilken2023dynamical}, with a measured value $\gamma \approx 0.58$, about 40\% larger than the theoretical prediction. 

Our data are shown in Fig.~\ref{fig:jamming}(c). They perfectly obey the power law in Eq.~(\ref{eq:zh}) over many decades in $h$, with the predicted mean-field value for the exponent $\gamma$, and in agreement with earlier numerics using energy minimization. The disagreement noted in Ref.~\cite{wilken2023dynamical} therefore does not stem from the dominance of random organization dynamics (linked to conserved directed percolation criticality) over jamming. Rather, the overestimation of the exponent $\gamma$ was due to numerical measurements performed too far from the jamming critical point. The data in Ref.~\cite{wilken2023dynamical} cover about 1 decade in $h$ near $h=10^{-3}$, as opposed to the 8 decades revealed in Fig.~\ref{fig:jamming}(c). In addition, we have numerically verified that removing the center of mass conservation from the dynamics does not alter this conclusion.

Our interpretation is that the critical behavior controlled by conserved directed percolation concerns physical observables that have no connection to the jamming criticality and related observables. The two distinct lower critical dimensions ($d_l=4$ for conserved directed percolation and $d_l=2$ for jamming) can thus coexist in the same system, because they concern distinct sets of physical observables that do not interfere.

\section{Hyperuniform density fluctuations in various phases}

\label{sec:hyperuniformity}

\subsection{Hyperuniform density fluctuations in size disperse systems}

The non-equilibrium dynamics of random organization models are known to produce hyperuniform configurations in which density fluctuations are suppressed at large lengthscales~\cite{hexner2015hyperuniformity,hexner2017noise,Lei2025nonequilibrium}. In mono-component systems, hyperuniformity is quantified by the scaling of the structure factor, Eq.~\eqref{eq:soq}. When density fluctuations are suppressed, one observes 
\begin{equation} 
S(q \to 0) \sim q^\alpha, 
\end{equation}
with $\alpha>0$ an exponent characterizing various forms of  hyperuniformity~\cite{torquato2018hyperuniform}. 

When size-disperse mixtures are studied, however, density and composition fluctuations are coupled and other metrics are needed~\cite{berthier2011suppressed}. For the binary mixture under study, it is convenient to study the $q$-dependent compressibility~\cite{berthier2011suppressed},
\beq
\chi(q)=\frac{S_{{AA}}(q)S_{{BB}}(q)-S_{{AB}}^2(q)}{x_{{A}}^2S_{{BB}}(q)+x_{{B}}^2S_{{AA}}(q)-2x_{A}x_{{B}}S_{{AB}}(q)},
\label{eq:chiq}
\eeq
where 
\beq
S_{\alpha\beta}(q)=\frac{1}{N} \left\langle \sum_{i\in \alpha} \sum_{j \in \beta} \mathrm e^{\mathrm i {\bf  q} \cdot ({\bf r}_i- {\bf r}_j)} \right\rangle 
\eeq
are the partial structure factor for species $\alpha, \beta = A, B$, and $x_\alpha$ denotes the concentration of species $\alpha$. It is clear that $\chi(q)$ becomes equal to $S(q)$ in the limit where one component appears with a vanishing concentration. 

Exactly at the absorbing transition critical point, hyperuniform configurations with a non-trivial exponent $\alpha$ are dynamically produced~\cite{hexner2015hyperuniformity}. The exact value of $\alpha$ across models and dimensions was studied using various renormalization group techniques~\cite{wiese2024hyperuniformity, ma2025hyperuniformity}. For our model in $d=2$ earlier studies have reported $\alpha \approx 0.45$~\cite{hexner2015hyperuniformity}. 

In the presence of center of mass conservation, the entire active phase is characterized by hyperuniformity with an exponent $\alpha=2$. This was observed in dilute~\cite{hexner2017noise,lei2019hydrodynamics,tjhung2015hyperuniform} and crystalline~\cite{activeXtal2023,maire2024enhancing} environments. In that case, one expects a critical crossover between $q^2$ and $q^\alpha$ regimes as $\phi_c$ is approached from the active phase. 

In the rest of this section, we study how hyperuniformity manifests itself in the three important parts of the phase diagram: active liquid, active glass, and at the jamming critical point. 

\subsection{Hyperuniformity in active liquid}

\begin{figure}
\includegraphics[width=\columnwidth,clip=true]{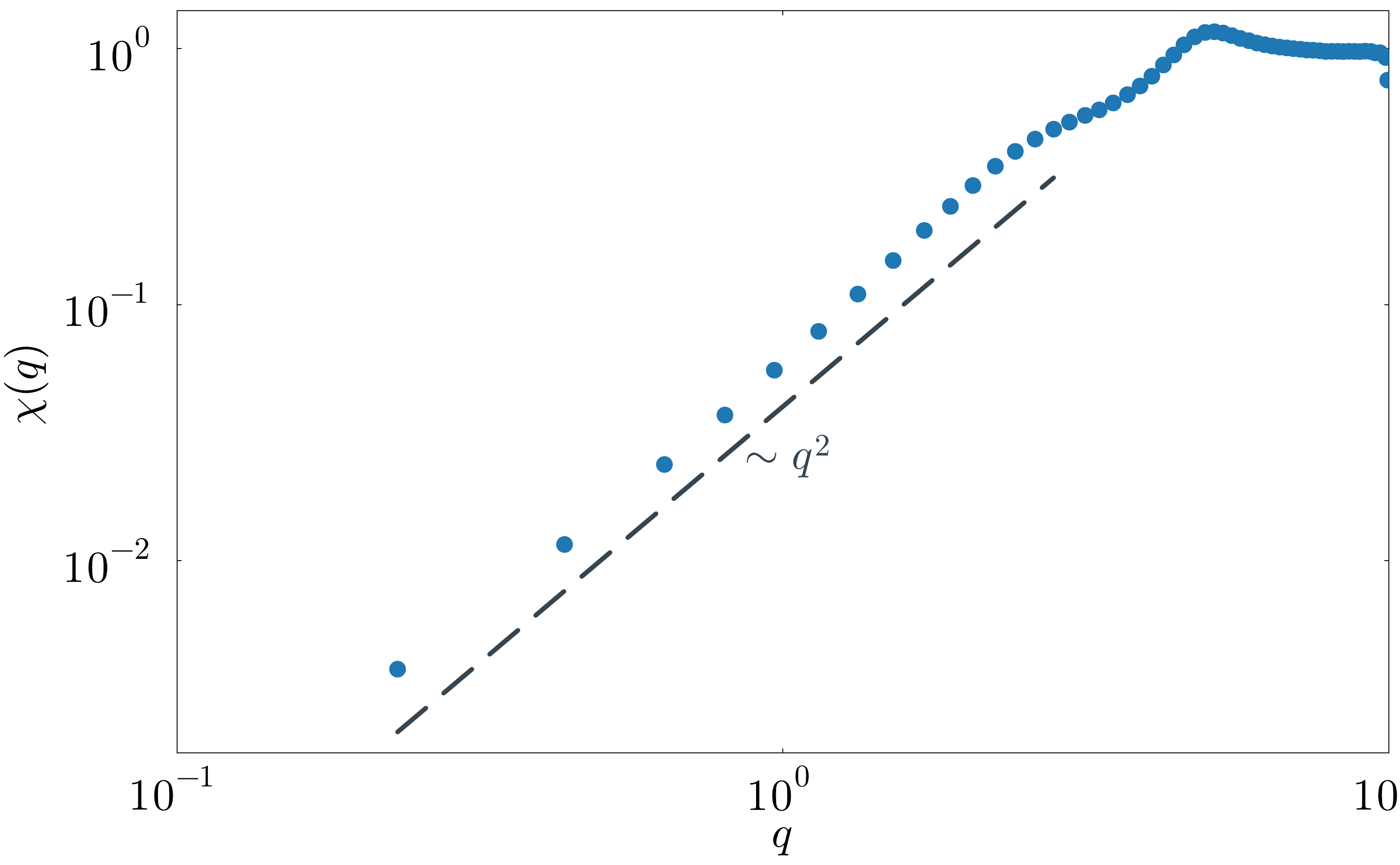} 
\caption{Compressibility in active diffusive liquid measured at $(\phi=0.65,\epsilon=0.6)$, with quadratic hyperuniform scaling at low $q$ (dashed line).} 
\label{fig:liquid_hyperuniformity} 
\end{figure}

In the active liquid, where dynamics is diffusive and the system thus resembles earlier studies at lower density, we expect the quadratic behavior observed for $S(q)$ to extend to the compressibility $\chi(q)$.

To confirm this expectation, we measure $\chi(q)$ in the steady state of the liquid phase for systems of $N=10^3$ particles, deep in the active liquid at $(\phi=0.65,\,\epsilon=0.6)$. The result is shown in Fig.~\ref{fig:liquid_hyperuniformity}, and confirms the expected quadratic scaling $\chi(q) \sim q^2$ discussed and reported earlier~\cite{hexner2017noise, Lei2025nonequilibrium}.

\subsection{Hyperuniformity in active glass: suppressed phonons and non-hyperuniform backbone}

The physics of density fluctuations is radically different in the glass phase, where the system is kinetically arrested and particle do not diffuse. Since the activity $\langle f \rangle$ remains large, particles constantly undergo pairwise collisions and perform random displacements of small amplitude. As a result, particles perform constrained localized motion around well-defined averaged positions. This is reminiscent of the physics of the corresponding non-equilibrium active crystals~\cite{activeXtal2023} with the important difference that here the averaged positions do not form a periodic lattice, but define instead an aperiodic, amorphous structure.  

Just as for the crystal case, it is convenient to decompose the instantaneous particle positions into two components: 
\begin{equation}
{\bf r}_i (t) = \overline{\bf r}_i + \delta {\bf r}_i (t),  
\label{eq:decomposition_r}
\end{equation}
where the overline describes a time average in a particular glass configuration. This decomposition naturally extends to the compressibility~\cite{ikedaThermalFluctuationsMechanical2015}.  
\begin{equation}
\chi(q) = \chi_0(q) + \chi_\delta(q), 
\label{eq:decomposition}
\end{equation}
where the backbone contribution $\chi_0(q)$ is computed from Eq.~(\ref{eq:chiq}) but using the time-averaged positions $\overline {\bf r}_i$. The fluctuating part $\chi_\delta(q)$ is then defined by difference,  $\chi_\delta (q) \equiv \chi(q) -\chi_0(q)$. \rev{The decomposition in Eq.~(\ref{eq:decomposition_r}) is well-defined when there is a clear separation of timescales between the relaxation of density fluctuations (first term) and the exploration of the local cage (second term). This is not possible in the diffusing liquid phase, nor close to the transition between liquid and glass. However, it is valid in the glass phase where aging effects are unable to relax density fluctuations on large lengthscales, corresponding to the low-$q$ behavior of $\chi_0(q)$.}

\begin{figure}
\includegraphics[width=\columnwidth,clip=true]{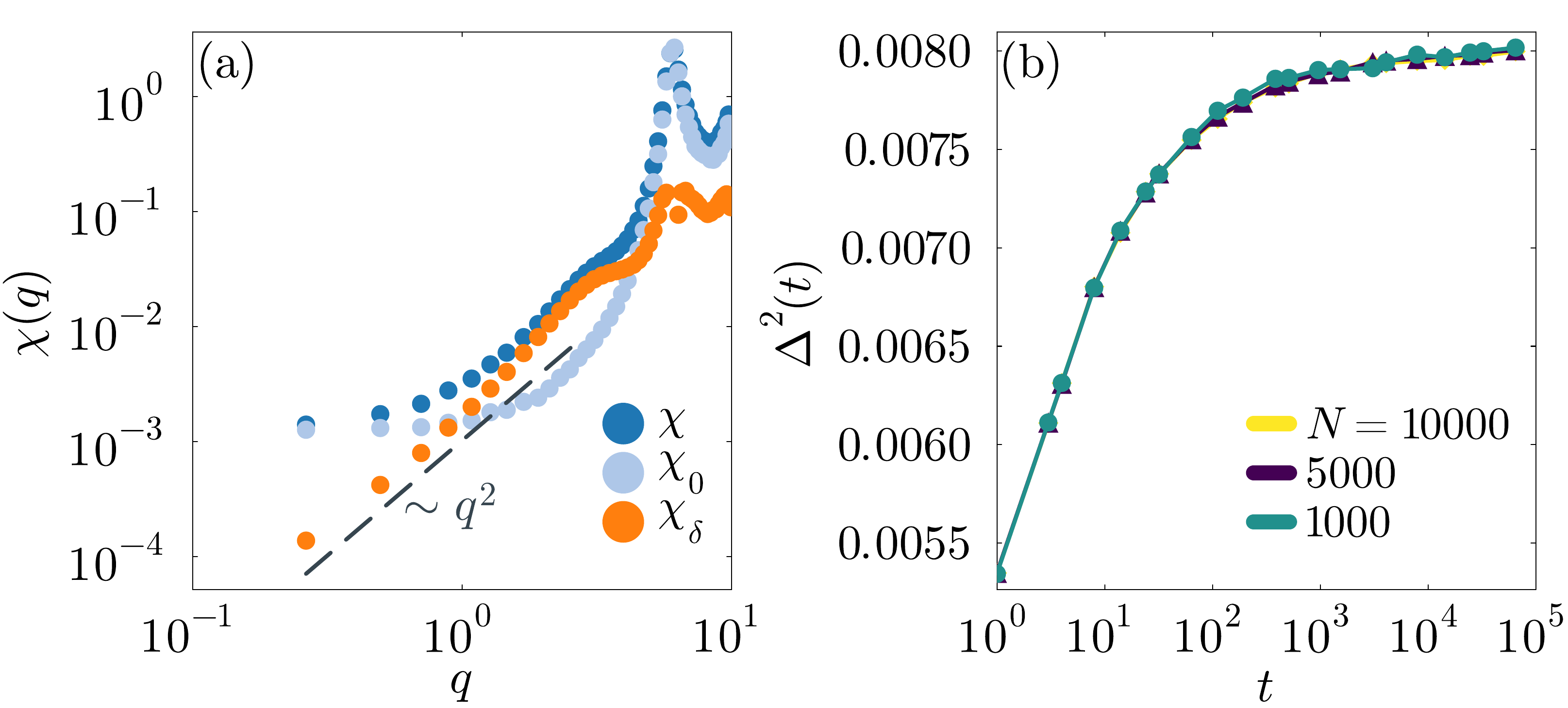} 
\caption{(a) Decomposition of the compressibility $\chi(q)$ along time-averaged aperiodic positions ($\chi_0(q)$) and instantaneous ($\chi_\delta(q)$) fluctuations for an active glass at $(\phi=0.844,\epsilon=0.1)$. The quadratic scaling $\chi_\delta(q) \sim q^2$ holds for fluctuations, whereas the backbone and total contributions reach a plateau in the $q \to 0$ limit. This reflects non-phononic vibrations of particles around averaged positions that form a non-hyperuniform amorphous structure.
(b) Mean squared displacement $\Delta^2(t)$ for different system sizes.
The plateau height $\Delta^2(\infty)$ is independent of $N$, confirming that the center-of-mass conservation regularizes the vibrational spectrum and suppresses the logarithmic divergence typical of equilibrium two-dimensional solids.
} 
\label{fig:glass_hyperuniformity} 
\end{figure}

These various contributions are compared in Fig.~\ref{fig:glass_hyperuniformity}(a) for an active glass state with $N=10^3$ particles at $(\phi=0.844, \epsilon=0.1)$. We used 24 independent glass configurations, each being followed over $10^6$ steps to correctly record time-averaged positions. The total compressibility $\chi(q)$ plateaus to a finite value as $q \to 0$. A similar behavior is observed for the backbone contribution $\chi_0(q)$, and both functions become equal at low $q$ where $\chi_\delta(q)$ becomes negligible. This indicates that the frozen backbone of the active glass is disordered and not hyperuniform, just as conventional thermal glasses prepared by cooling.  

The non-equilibrium nature of the model is only perceived in the fluctuating contribution that inherits the quadratic scaling, $\chi_\delta(q) \sim q^2$, observed for the total compressibility $\chi(q)$ in the liquid. Physically, the glass transition essentially freezes the particles not very far from their initial positions, and the system then performs non-thermal vibrational motion near a frozen backbone. There is no deep physical reason for this arrested backbone to be hyperuniform. By contrast, the fluctuating part of the particle positions $\delta {\bf r}_i$ is controlled by the random organization dynamics that suppresses phonons at low $q$~\cite{activeXtal2023}, thus leading to the observed $\chi_\delta(q) \sim q^2$ scaling.

An essential difference between periodic and non-periodic situations is that the periodic lattice gives no contribution to $\chi(q)$ at low $q$ (in Fourier space $\chi_0$ is a collection of discrete Bragg peaks), and the decomposition in Eq.~(\ref{eq:decomposition}) is not needed to reveal the non-phononic nature of the vibrations. As a result, $\chi(q)$ displays hyperuniform behavior in the crystal~\cite{activeXtal2023}, because it is dominated by $\chi_\delta(q)$ at low $q$ where $\chi_0 =0$. 

In two-dimensional particle systems at equilibrium, the existence of thermal phonons is directly revealed via the existence of Mermin-Wagner fluctuations~\cite{MerminWagner} that ultimately prevent the existence of long-range order~\cite{mermin1968crystalline}. In practice, Mermin-Wagner fluctuations can be quantitatively revealed by a strong system size dependence of the mean-squared displacement $\Delta^2(t)$ that reaches, at long times $t \to \infty$, a plateau that diverges logarithmically with the system size $N$. These long-wavelength fluctuations perturb the glassy dynamics of two-dimensional thermal glasses~\cite{flenner2015fundamental}, but they were shown to be suppressed in two-dimensional crystals evolving under random organization dynamics with center-of-mass conservation~\cite{activeXtal2023,maire2024enhancing}.   

We show in Fig.~\ref{fig:glass_hyperuniformity}(b) that the latter conclusion also applies to the active, amorphous glass state. The mean-squared displacements measured for three different system sizes perfectly overlap at all times, including in the final plateau regime. We verified numerically that removing the conservation law re-introduces for the same state point a pronounced system size dependence, thus validating our conclusions that phonons are suppressed at large scale in the active glass in the present model. However, these anomalous (hyperuniform) vibrational modes live on top of an otherwise non-hyperuniform amorphous background that is not controlled by random organization dynamics.  

\subsection{Non-universal hyperuniformity at the jamming transition}

The characterization of density fluctuations precisely at the jamming transition benefits from the above separation between the frozen backbone and the instantaneous fluctuations. A similar distinction in fact holds for thermal packings of soft spheres~\cite{ikedaThermalFluctuationsMechanical2015}, where conventional phonon modes control instantaneous fluctuations over a frozen amorphous backbone.

In the $(\phi,\epsilon)$ phase diagram, the activity vanishes along the critical line $\phi_c(\epsilon)$. This line can be approached for instance by decreasing $\phi$ at constant $\epsilon$. When $\epsilon$ is large enough, the approach to $\phi_c(\epsilon)$ is from the diffusive fluid. In that case, the structure factor was shown to obey $S(q) \sim q^\alpha$ with $\alpha \approx 0.45$~\cite{hexner2015hyperuniformity}, a universal exponent related to the conserved directed percolation universality class~\cite{wiese2024hyperuniformity}. 

We expect a different physics in the small $\epsilon$ limit as the system now approaches the critical absorbing transition from the active glass state, and it can therefore retain memory of its past history. If there is no rearrangement in the glass phase, the backbone structure is essentially unaffected by the approach to the critical absorbing line $\phi_c$, and the backbone structure at the critical point remains protocol-dependent. This directly proves that the conserved directed percolation universality class, while controlling the criticality and physics of instantaneous fluctuations, does not control the large-scale density fluctuations that have been frozen by quenching the system into the active glass phase. 

We pay special attention to the $\epsilon \to 0$ limit of the absorbing line that corresponds to a jamming transition. While conserved directed percolation does not impose its hyperuniform behavior to $\chi(q)$ at any $\phi_c(\epsilon)$, jamming physics could instead be responsible for the emergence of hyperuniformity at $\phi_J$. On this point, the literature is considerably divided about the emergence of perfect or imperfect hyperuniformity at jamming, with a broad diversity of behaviors reported across various computational approaches, models and dimensions~\cite{donev2005unexpected,ikedaThermalFluctuationsMechanical2015,ikeda2017large,maher2023hyperuniformity,maher2024hyperuniformity,miyazaki2025hyperuniformity}.  

To test our conjecture, we follow a similar annealing protocol as above to closely approach the $\phi_J$ limit using random organization dynamics. To gather sufficient statistics, we quench 64 independent configurations to the same state point $(\phi=0.84,\epsilon=0.01)$ starting from two independent sets of initial conditions to also test universality. We choose initial conditions from hard disk equilibrium configurations at $Z_0=0$ (random initial conditions) and from an equilibrated fluid state at $Z_0=24$ (corresponding to $\phi_0 \approx 0.787$, in the equilibrium fluid phase). We then slowly anneal the value of $\epsilon$ along the $\phi_c$ line, as explained in Sec.~\ref{sec:jline}. We use the parameters $\lambda = 3.7 \times 10^{-7}$ and $T = 5 \times 10^7$ to reach $\phi_J$. We checked that the obtained compressibility no longer evolves, within statistical accuracy, if we decrease $\lambda$ further. 

\begin{figure}
\includegraphics[width=\columnwidth,clip=true]{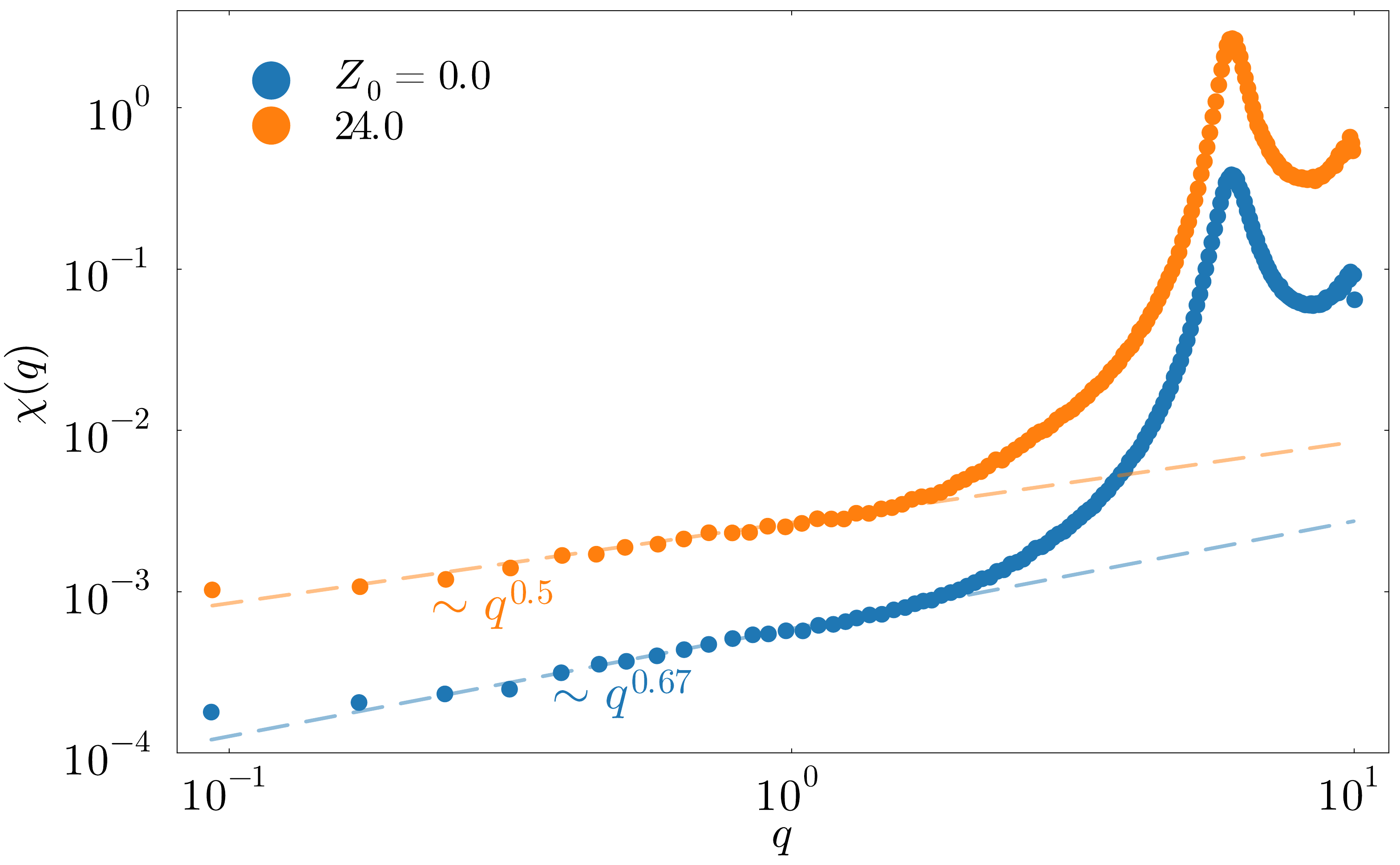} 
\caption{Compressibility $\chi(q)$ for jammed configurations reached via slow annealing starting from two different preparation pressures: random initial conditions ($Z_0=0$, blue) and from finite pressure ($Z_0=24$, orange, shifted vertically for clarity). The different exponents demonstrate that the large-scale structure of the jammed state is not universal.} 
\label{fig:jamming_hyperuniformity} 
\end{figure}

\rev{The results in Fig.~\ref{fig:jamming_hyperuniformity} confirm that nearly hyperuniform behavior emerges at $\phi_J$ for the total compressibility $\chi(q)$ (recall that $\chi_\delta(q) \to 0$ at jamming), which is well described by a non-trivial power law, $\chi(q) \sim q^\alpha$. It is difficult to decide whether this power law behavior extends to $q \to 0$ or is instead only observed over a finite range of wave vectors, here $q \sim 0.01 - 3$. This would require slowly preparing much larger packings at the jamming transition, which is numerically very demanding.  
}

A first notable observation is that the scaling exponent $\alpha \approx 0.67$ for $Z_0=0$ is consistent with numerical findings for a similar bidisperse mixture of soft sphere prepared via energy-minimization compression~\cite{miyazaki2025hyperuniformity}. This agreement confirms our hypothesis that it is jamming physics (rather than conserved directed percolation and random organization) that controls the compressibility when jamming is approached, even when random organization dynamics is employed. This conclusion is further supported by the fact that the value $\alpha \approx 0.67$ is far above the value expected value for conserved directed percolation ($\alpha=0.45$). 

A second noticeable result is that the measured exponent $\alpha$ is non-universal, as a continuous range of values is obtained for a different set of initial conditions, for instance using a finite $Z_0>0$. The non-universal nature of hyperuniform signatures at jamming was noted before in the context of compression algorithms~\cite{ozawa2017exploring}, and this seems to be generically true. The general trend is that the hyperuniform suppression of density fluctuations obtained for nearly random initial conditions approaching jamming, becomes less prominent when starting from denser initial conditions. In our case, the effective $\alpha$ exponent indeed decreases for larger $Z_0$.  

Our broad conclusion is that the large-scale structure at jamming of packings at the jamming transition is not controlled by the physics of random organization. In particular, the conserved directed percolation exponent $\alpha$ is not relevant to describe jamming. We also suggest that predicting a hyperuniformity exponent at the jamming transition is again not a well-posed question, as $\alpha$ is protocol-dependent, and therefore non-universal, just as the value of $\phi_J$ itself. The non-universality of $\alpha$ at jamming is consistent with the wide diversity of reported exponents in the literature across a diversity of preparation protocols. We suggest that this diversity is not due to the quality of the preparation of jammed packings, but rather illustrates the non-universal nature of hyperuniformity of jammed packings. 

\section{Conclusion}

\label{sec:conclusion}

Let us now summarize our main findings, and discuss our central conclusions. The random organization model studied in this work possesses multiple non-equilibrium features: it undergoes distinct glass and jamming transitions, as well as a critical line towards absorbing states. \rev{This competition can exist because the random displacements performed at each step are biased. For isotropic displacements, the transition line stops at much smaller density, and the physics is less rich~\cite{tjhung2015hyperuniform}.} The interplay between the criticality and heterogeneity associated to each of those aspects makes the physics quite rich, and multiple observables were shown to display non-trivial behavior that are either universal or protocol-dependent. Our goal has been to clarify these observations, and resolve conflicting literature results.

Our central finding is the observation that the active phase above the absorbing transition line becomes glassy when $\epsilon$ is small because the corresponding packing fraction $\phi>\phi_c(\epsilon)$ becomes large enough for crowding effects to become dominant. As a result, the active glass phase retains a permanent memory of its preparation history because particles can only move a finite distance from their initial positions. The fundamental consequence is that most properties of the active glass phase are then potentially protocol-dependent, and therefore possibly non-universal. 

First of all, we showed that the location of the absorbing line $\phi_c(\epsilon)$ is protocol-dependent, including its $\epsilon \to 0$ limit that defines a jamming transition. These findings in fact parallel the conclusions made earlier in the context of jamming studies, where the dependence on the protocol is also unavoidable~\cite{chaudhuri2010jamming}. All these results show that predicting the location of the jamming transition is meaningless without a prescription for the dynamic protocol used to produced jammed packings. 

This suggests that random organization dynamics cannot be used to propose a better definition of random close packing compared to earlier approaches, such as energy-minimization or compression protocols. Random organization is in fact quite similar to these earlier studies. We also suggested that the definition of a jamming transition as the limit $\phi_c(\epsilon \to 0)$~\cite{wilken2021random} is in fact nearly equivalent to the definition of the point-$J$ of O'Hern and coworkers~\cite{ohern2002random}. In both cases, there exists a continuous line of jamming transitions, sometimes called the `$J$-line'~\cite{chaudhuri2010jamming}, as reported in numerous studies~\cite{berthier2009glass,chaudhuri2010jamming,ozawa2012jamming,ozawa2017exploring,berthier2024Monte}. 

Although the location of these phase transitions is protocol-dependent, we showed that several features remain universal. For instance, the criticality of the absorbing phase transition (such as the evolution of activity, its spatial and temporal correlations) does not depend on the protocol, and thus universality is preserved. There have been reports of deviations from the conserved directed percolation universality class for both ordered and disordered states~\cite{ghosh2022coupled,wang2025anomalous}. We suggest that this only arises when the system is quenched from fully random initial conditions. In that case, the slow aging of the global structure competes with the approach to the absorbing line. We have shown that by first preparing a system in the active phase before gradually approaching the absorbing line removes the structural aging contribution, and conventional criticality is then retrieved. We reached the same conclusion before in the case where the active phase displays periodic order~\cite{activeXtal2023}. 

The observation that a jamming transition can be defined using random organization dynamics raises the question of how these two types of criticality compete or coexist in the $\epsilon \to 0$ limit. We have shown that conserved directed percolation only controls the behavior of observables related to the absorbing phase transition (the activity and its space-time fluctuations), but that jamming criticality is insensitive to the dynamic protocol used to create the packings. In particular, we found that the critical exponent controlling the gap distribution measured with random organization dynamics is the same as that measured in protocols using energy-minimization, and showed that the measurements reported in Ref.~\cite{wilken2023dynamical} were taken too far from the critical jamming point. This also invalidates the suggestion in Ref.~\cite{wilken2023dynamical} that the lower critical dimension of conserved directed ($d_l=4$) percolation influences the critical properties of jamming. We suggest that the coexistence of distinct lower critical dimensions in the same system is possible, because they concern different set of observables.    

Another connection proposed between random organization dynamics and jamming is related to the emergence of hyperuniformity. With center of mass conservation, random organization produces hyperuniform configurations both in the active phase (with $\chi(q) \sim q^2$) and at the critical point (where $\chi(q) \sim q^\alpha$, with $\alpha$ a non-trivial universal exponent). We demonstrated that when approaching larger densities, the emergence of a glass phase completely changes this picture. In the non-diffusive glass phase, random organization controls the fluctuating part of the particle motion, that takes place on an otherwise frozen density pattern. We showed that hyperuniformity applies to the fluctuations ($\chi_\delta \sim q^2$) but not to the frozen backbone whose structure is history dependent and not hyperuniform ($\chi_0(q) \sim \chi(q) \sim {\rm const}$). The active glass phase is not hyperuniform, but shows instead interesting non-phononic vibrational motion. In the limit of $\epsilon \to 0$, when jamming is approached, the backbone contribution again shows signs of hyperuniformity, with a behavior that is quantitatively consistent with earlier observations in soft sphere packings. We demonstrated that the hyperuniformity observed at jamming hyperuniformity is protocol-dependent and is therefore not universal. This shows in particular that conserved directed percolation cannot be used to infer the large-scale structure of packings at jamming. The protocol-dependence exposed here confirms earlier studies, and rationalizes the diversity of hyperuniformity exponents reported in sphere packings at jamming.

The interplay between crowding and non-equilibrium random organization dynamics provides an interesting perspective on both problems. On the one hand, random organization provides a new way to approach jamming and this has raised new questions. On the other hand, this allows to test what aspect of the physics is simply due to crowding or is instead controlled by a specific microscopic dynamics. It also helps to disentangle universal from non-universal aspects, as well as test the validity and pertinence of existing theoretical approaches and predictions.    

\section*{Acknowledgments}

We thank D. Levine and S. Wilken for useful exchanges about random organization models. We also thank E. Corwin, P. Charbonneau, and K. Wiese for discussions about these results. L.G. acknowledges co-funding from the European Union - Next Generation EU. L.B. acknowledges the support of the French Agence Nationale de la Recherche (ANR), under grants ANR-20-CE30-0031 (project THEMA) and ANR-24-CE30-0442 (project GLASSGO).

\section*{Data Availability}

\rev{The data that support the findings of this article are
openly available~\cite{zenododataset}.}

\bibliography{main.bib}

\end{document}